\documentstyle[12pt,psfig]{article}
\def\al{\alpha}                                  
\def\b{\beta}                                   
\def\g{\gamma}                                  
                                  
\def\ep{\epsilon}             
                                   
\def\al{\alpha}                                  
\def\b{\beta}                                   
\def\g{\gamma}                                  
                                  
\def\ep{\epsilon}             
                                   
\def\p{\pi}

\def\t{\tau}

\def\f{\frac}
\def\pr{\partial}
\def\eq{equation}

\begin{document}

\vspace*{2mm} 
\begin{center}
{\Large
Elimination of Chaos in Multimode, Intracavity-doubled}
\end{center}
\begin{center}
{\Large Lasers in the Presence of Spatial Hole-burning.}
\end{center}
 
\vspace{3mm} 

\begin{center}
{\large
{\it Monika E. Pietrzyk}\\
Faculty of Physics, Warsaw University of Technology\\
{\sc Warsaw, Poland}}\\
\end{center}

\begin{center}
{\large
{\it Miltcho B. Danailov}\\
{Laser Laboratory, Sincrotrone}\\
{\sc Trieste, Italy}}\\ 
\end{center}

\vspace{3mm}

\begin{abstract}
In this paper possibilities of a stabilization of large amplitude
fluctuations in an intracavity-doubled solid-state laser are studied.
The modification of the cross-sa\-tu\-ra\-tion coefficient by the effect of 
spatial hole-burning is taken into account. The stabilization of the 
laser radiation by an increase of the number of modes, as proposed in 
[James et al., 1990b, Magni et al., 1993],
is analyzed. It is found that when the 
cross-saturation coefficient is modulated by the spatial hole-burning 
the stabilization is not always possible. We propose a new way of 
obtaining a stable steady-state configuration based on an increase of 
the strength of nonlinearity, which leads to a strong  
cancellation of modes, so that during the evolution all of the 
modes, but a single one, are canceled. Such a steady-state solution 
is found to be stable with respect to small perturbations. 
\end{abstract}

\vspace{1cm}

Short title: 
Elimination of Chaos in Multimode, Intracavity-doubled...

\pagebreak 

\section {Introduction}

Solid-state lasers containing frequency-doubling crystals are 
efficient and compact sources of co\-he\-rent visible optical radiation. 
Unfortunately, when they operate in multimode regime, one observes
irregular fluctuations of the output intensity. This behavior, referred
to as the green problem, has been reported for the first time by Baer 
[Baer, 1986]. He found that these instabilities 
arise from a coupling between longitudinal modes of the laser due 
to sum-frequency generation. In particular, when such a laser 
operates in a single
longitudinal mode, its output is stable 
[Kennedy \& Barry, 1974]. 
In the case of two oscillating longitudinal modes, the output intensity 
is stable only for small values of nonlinearity, 
otherwise both modes tend to pulse on and off out of phase 
[Baer, 1986].
When the number of lasing modes is larger than two, 
the laser can exhibit, depending on the parameters describing it, 
various types of behaviour such as: 
antiphase dynamics 
[James et al., 1990b, Wiesenfeld et al.,1990, James et al., 1990a, 
Roy et al.,  Bracikowski \& Roy, 1991, Mandel \& Wang, 1994, 
Wang et al., 1995, Otsuka et al., 1997], 
clustering 
[Baer, 1986],
grouping 
[Otsuka et al., 1997],
and chaotic dynamics 
[James et al., 1990a, Roy et al., Bracikowski \& Roy, 1991, 
Liu et al., 1997].

One of the motivations of the large number of papers 
devoted to intracavity-doubled solid state lasers and published
during the last decade has been the need 
to find a configuration and the range of parameters describing the lasers,
which could guarantee a stability of their outputs.
Several ways to reach this goal have been found, for example, 
an introduction of a quarter-wave plate into the laser cavity, 
a proper alignment of the angle between birefringent 
axes of the active medium and the nonlinear crystal 
[James et al., 1990a, James et al., 1990b, Roy et al.],
a placing of a tilted mirror inside the cavity 
[Ustyugov  et al., 1997],
or an utilization of the effect of sum-frequency-mixing 
[Falter et al., 1997, Danailov \& Apai 1994].
Moreover, it has been predicted 
analytically 
[James et al., 1990b]
and confirmed experimentally 
[Magni et al., 1993]
that when the number of oscillating longitudinal modes is set
to be sufficiently large (for example, by increasing the length of 
the laser cavity) the output of the laser becomes stable. Besides, 
the green problem can also be avoided in the case of a single mode 
operation of the laser, which can be obtained, for example, by use of 
an intracavity etalon 
[Baer, 1986]
or a birefringent filter 
[Nagai et al., 1992, Fan, 1991]

The main goal of this paper is to continue the investigations of the
green problem. The fact that the cross-saturation coefficient is modulated 
by the spatial hole-burning effect is taken into account. 
Firstly, the stabilization of the 
laser radiation by an increase of the number of longitudinal modes, as
proposed in 
[James et al., 1990b, Magni et al., 1993]
is analyzed. The results presented 
in 
[James et al., 1990b, Magni et al., 1993],
where the cross-saturation coefficient
was assummed to be constant for all modes, are compared with the 
numerical data. It is shown that the theoretically 
obtained 
[Wang \& Mandel, 1993]
linear dependence of the minimal number 
of modes necassary for stabilization the laser output on the strength 
of nonlinearity agrees with the numerical solutions only 
in the case of sufficiently small nonlinearity. 
For larger values of  nonlinearity, due to the cancellation of modes
during the evolution, the minimal number of modes obtained 
by the numerical simulations is larger than the number which follows 
from the theoretical predictions. 
For very large nonlinearity this cancellation is so 
strong that only a few modes survive (even when there are initially 250 
oscillating modes in the laser cavity). Therefore, a large number of 
simultaneously oscillating modes necessary for stabilization 
of the laser output cannot be achieved. 
A similar situation takes place when the cross-saturation coefficient is 
modulated by the effect of spatial hole-burning, therefore
the stabilization of the laser output by an increase of the number of 
longitudinal modes is not always possible, as a result of a strong 
competition between modes and a cancellation of some of them during the 
evolution. However, the problem of the stabilization of the laser 
output can be solved in another way, namely, by an increase of the strength 
of nonlinearity, which leads to very strong competition
between the modes, so that during the evolution all of them, 
but a single one, are canceled. 
As a consequence, a steady-state solution, which is stable against small 
perturbations, arises. This is the stabilization mechanism proposed
in the present paper. It is valid in the case of constant 
cross-saturation coefficient, as well as in the case when the spatial 
hole-burning is taken into account.

\section{Basic Equations}

The analysis presented in this paper is based on the Baer-type rate 
equations 
[Baer, 1986]
extended by Roy, Bracikowski and James to the case
when the effect of spatial hole-burning is taken into account
[Roy et al., Bracikowski \& Roy, 1991]
\footnote{Although Roy et al. have written down an explicit form for the
cross-saturation coefficient, they were using in their analysis
the approximation that the cross-saturation coefficient is constant
for all modes.}. These equations have the following form:
\large
$$
\left\{
\begin{array}{c}
\t_{c1} \f{\pr I(p,t)}{\pr t} = I(p,t) \left( -\al_p + G(p,t) 
- \ep I(p,t) -2 \ep\sum_{q \neq p} I(q,t) \right), \hspace{2cm} (1a)\\
\\
\t_f \f{\pr G(p,t)}{\pr t} = G_{ap} -G(p,t) \left(1+ \b(p,p) I(p,t) 
+ \sum_{q \neq p} \b(p,q) I(q,t) \right), \hspace{1cm} (1b)
\end{array}
 \right.$$

\normalsize

\addtocounter{equation}{1}

\begin{\eq}
\b(p,q) = \b_0 \f{\int_{z_0}^{z_0+l} \left(1-cos(2k_pz)\right) 
\left(1-cos(2k_qz)\right)dz}
{\int_{z_0}^{z_0+l} \left(1-cos(2k_pz)\right) \left(1-cos(2k_pz)\right)dz} 
\approx \b_0 \left| \f{z+ \f{sin(2(k_p-k_q)z)}{4(k_p-k_q)}}{\f 32 z} 
\right|_{z_0}^{z_0+l}, 
\label{cross}
\end{\eq}
\[ k_p=\f {\p p} L, \hspace{5mm} p,q = 1,...,N, \]
where $N$ is the number of longitudinal modes;
$\t_c$ and $\t_f$ are, respectivelly, the cavity round trip time and the 
fluorescence lifetime; $I(p,t)$ and $G(p,t)$ are,
respectively, the intensity and gain of the $p$-th longitudinal
mode; $\al_p$ is the cavity loss parameter for the $p$-th mode; $\g_p$ 
is the small signal gain; $\b(p,p)$ and $\b(p,q)$ describe, 
respectivelly, self-saturation of the $p$-mode and cross-saturation between 
two modes,  $p$ and $q$; $\b_0=0.06$ is the scaling parameter; $l$ 
is the length of the gain medium; $L$ is the length of the laser cavity,
$z_0$ is the distance of the gain medium from the first cavity mirror,
(as shown in Fig. 1); $k_p = \f{2 \pi}{\lambda_p}$ 
is the wavevector of the longitudinal cavity mode $p$ 
with the wavelengths $\lambda_p = \f{2 L}{p}$.
Note that even through in Eq. (\ref{cross}) three parameters: $L$, $l$, 
and $z_0$ are present, the modified cross-saturation 
coefficient depends only on two 
rescaling parameters, $\f lL$ and $\f {z_0} L$.
The parameter $\ep$ is a nonlinear coefficient 
which describes the conversion efficiency
of the fundamental intensity into the doubled intensity. 
The terms $\ep I(p,t)^2$ and $\ep I(p,t) I(q,t)$ in Eq. (1 a)  
account for the loss in the intensity of the fundamental 
frequency through second harmonic generation and through sum-frequency 
generation, respectivelly. 

The set of equations (1 a,b) has a rather complicated structure and it is 
not possible to solve it analytically. Therefore, the results presented
in this paper has been obtained with the aid of
the numerical method of Runge-Kutta.

\section 
{Simplified Model - No Spatial-Hole Buring Effect}

In this section the approximation that the cross-saturation coefficient 
is constant for all modes: $\b(p,q)=\b(\bar p, \bar q) = \f 2 3$, is 
used. It is assumed that losses and small signal gains are the same 
for all modes, i.e. $\al_p=\al_q=\al, \, G_{ap}=G_{aq}=G_{a}$, where
$p,q = 1,...,N$. Other parameters describing the system have been chosen 
as follows: $\t_{c1}=10 \, [ns], \t_f=0.24 \, [ms], \al=0.015, \g=0.12, 
\b_0=0.06$. The number of longitudinal modes, $N=1,...,250$,
and the strength of nonlinearity, $\ep = 10^{-7} \div 10^{-3}$,
are not fixed and they are varied in the analysis. 

As the first step let us consider the dependence of the
laser output on the number of longitudinal modes initially excided in 
the laser cavity. From the numerical simulations it follows that
for very small nonlinearity, $\ep \approx 0$, the laser output 
is stable, even for a large number of longitudinal modes. In the
case of larger nonlinearity, $\ep=1.0 \times 10^{-4}$, as it can be 
seen from  Fig. 2a, the behavior of the laser output is complicated
and already for three simultaneously oscillating longitudinal modes
the total intensity exhibits irregular oscillations. 
When the number of modes increases, amplitudes of those 
oscillations slowly increase (see Fig. 2b representing the results 
for 70 modes). Finally, when the number of modes is larger than the critical 
value, the total stabilization of the laser output occurs, as shown in 
Fig. 2c. 

However, for slightly higher than previously considered
value of nonlinearity, $\ep=1.2 \times 10^{-4}$,
we have not observed 
the stabilization of the laser output, even for very 
large number of longitudinal modes, $N=200$ (compare the total intensity 
of the laser output for $N=25$, $N=100$, and $N=200$ modes, shown,
respectivelly, on Figs. 2d, 2e, and 2f). This happens because
of cancelation of modes, as a result of a competition between them,
which takes place during the evolution. 

The dependence of the minimal number of modes neceessary for stabilization 
of the laser output on the strength of nonlinearity is presented in 
Fig. 3. As it can be seen, for small 
nonlinearity, $\ep < 0.8 \times 10^{-4}$, 
there is a rather good agreement between 
theoretically  
[Wang \& Mandel, 1993]
and numerically obtained results. 
However, when nonlinearity increases, 
$0.8 \times 10^{-4} < \ep < 1.0 \times 10^{-4}$,
the minimal number of modes obtained in numerial simulations 
is larger than the number predicted theoretically. 
This discrepancy is the
result of the competition between modes, which leads to 
cancellation of some of them during the evolution. 
When $\ep > 1.2 \times 10^{-4}$ the cancellation is so big that 
after some time only few modes survive, thus a sufficiently large
number of modes necessary for stabilization of the laser cannot be
realised.

Concluding, depending on the relation between nonlinearity 
and the number of lon\-gi\-tu\-di\-nal modes, the dynamics of a 
multimode, intracavity-doubled, solid-state laser 
can be divided into four regions as illustrated in Fig. 5. 
In particular, when nonlinearity is small and the number of modes 
is larger than a critical value the output of the laser is stable.
Examples of such a behavior, which is labeled as Region I,  
are shown in Fig. 4a. With increasing nonlinearity 
the complexity of the behaviour of the 
laser increases and irregular fluctuations 
of the output intensity appear. This behavior, depicted in
Fig. 4b, corresponds to Region II. For even larger values of the
strength of nonlinarity the cancellation of modes starts to take place.
When nonlinearity exceedes the critical value $\ep_{crit1}$ 
the evolution of the output intensity leads to a cancellation of a large
number of modes, 
so that only a few of them survive. As s result, a stabilization of the
laser output cannot be obtained. This behavior, illustrated in Fig. 4c, 
is classified as Region III.
When nonlinearity is larger than the critical value $\ep_{crit2}$
there is such a large mutual cancellation of modes that after some time 
only one of them survives. Therefore, the steady-state solution, shown in 
Fig. 4d, which is stable against small perturbations, arises. 

This mechanism of stabilization, achieved by forcing the laser to operate 
in the one-mode regime, is similar to other approaches presented in the 
literature, where the stabilization is obtained by inserting into the laser
cavity an additional element like 
an etalon 
[Baer 1986]
or a birefringent crystal 
[Nagai et al., 1992, Fan, 1991].
However, the method proposed here seems to be a better solution, since
no additional element is involved.

\section{Inclusion of the Effect of Spatial Hole-Burning}

In this section the analysis presented in the previous section
is extended to take into account the effect of spatial hole-burning. 
The present analysis is based on the Baer-type rate equations, Eq. (1),  
modified by Bracikowski, Roy and James 
[Roy et al., Bracikowski \& Roy, 1991],
who have written down an explicit form for the 
cross-saturation coefficient, Eq. (2), but were not investigating
the influence of the spatial hole-burning on the laser dynamics. 
As far as the autors know, the analysis presented in this section 
is the first study of this issue. 

Similarly to the previous section, 
two ways of the laser stabilization are analyzed: 
by increasing the number of longitudinal modes and by increasing the 
strength of nonlinearity.  Parameters describing the laser cavity
are chosen to be as follows:
$L=1.2 \, [m],$ $\l=6 \, [mm],$ $z_0=12 \, [mm]$.

Our results show that when the spatial hole-burning effect 
is taken into account (i) the competition between the modes is much 
stronger than in the case of the cross-saturation coefficient 
constant for all modes (ii). Therefore, the minimal number of 
modes which are necessary to obtain the stabilization of the laser 
output is in the case (i) larger than in the case (ii). For example, 
for the strength of nonlinearity $\ep = 0.0001$ only 80 modes are 
necessary in the case (ii), as shown on Fig. 2c, while in the case (i) 
even 200 initially oscillating modes are not sufficient for the
stabilization of the laser output, as shown on Fig. 6c. 

Therefore, an alternative approach of the stabilization of the laser 
output based on an
increasing of the strength of nonlinearity, which has been proposed in the
previous section, can be examined. Indeed, 
from the results of the numerical simulations, displayed in Fig. 6f 
it follows that the stabilization of the laser output can be
obtained when the nonlinearity is larger than a critical value. 
This occurs because of an extensive cancellation of modes, due to which
the laser operates in a single longitudinal mode.

\section{Conclusions}

In this paper the
properties of the dynamics of multimode intracavity-doubled 
solid-state lasers have been studied. The infulence of 
spatial hole-burning on the cross-saturation coefficient
was taken into account. 
The system was described by the rate equations of Baer-type, 
which were solved numerically with the aid of the Runge-Kutta method.

Firstly, the case of the cross-saturation coefficient constant 
for all modes was considered. The minimal number of modes 
necassary for the stabilization of the laser output 
was defined and its dependence on the strength of 
nonlinearity was studied. It was shown that for small
nonlinearity this dependence is linear and agrees with 
the theoretical predictions. However, for larger values of nonlinearity 
the minimal number of modes was found to be larger than the number
obtained in theoretical considerations. As a reason of this discrepancy 
the effect of the competition between the modes and cancellation of some
of them during the evolution was given. It was also observed that 
with increasing of nonlinearity the competition between modes increases, 
so that for some values of nonlinearity the initially 
large number of oscillating modes is reduced to only a few of them. 
For much higher values of nonlinearity this competition is so strong that 
only a 
few modes survive. Therefore, a large number of simultaneously 
oscillating modes cannot be realised and the stabilization cannot be 
obtained in this way. However, we propose 
another method based on the increase of the
strength of nonlinearity. This stabilization occurs for
such a high value of nonlinearity, for which all modes,
besides a single one, are canceled during the evolution. 
In this case a steady-state solution, 
stable against small perturbations, arises.

A similar analysis was accomplished for the cross-saturation
coefficient modulated by the effect of spatial hole-burning.  
We have shown that in this case the stabilization of the laser output
by increase of the number of longitudinal modes is hardly possible. 
Therefore, an alternative approach of the stabilization based on an
increasing of the strength of nonlinearity have been examined and shown 
to be an appropriate solution. 

The method of a stabilization of a laser output, as proposed in our paper,    
is similar to other approaches presented in the literature where the 
stable output of the laser is achieved by forcing the laser to operate 
in a single-mode regime, 
for example, by inserting into the cavity an etalon or a birefringent 
crystal. However, the method proposed here offers a better solution, 
since no insertion of an additional element into the laser cavity 
is needed. 

\section{Acknowledgment}

The authors are grateful  
to Prof. M. Virasoro, the Director of the Abdus Salam
International Centre for Theoretical Physics, Trieste (Italy),
for granting this work and to Prof. G. Denardo for his encouragement; 
M. D. wishes to acknowledge the support of the International Centre 
for Science and High Technologies (I.C.S.), Trieste (Italy).


\pagebreak

\section{References}

\begin{itemize}

\item[]{Baer, T. [1986] 
{"Large-Amplitude Fluctuations Due to Longitudinal Mode 
Coupling in Diode-Pumped Intraavity-Doubled Nd:YAG lasers,"}
{\it J. Opt. Soc. Am. B} {\bf 3}, 1175.}

\item[]{C. Bracikowski, C. \&  Roy, R. [1991]
{"Chaos in a Multimode Solid-State laser System,"}
{\it Chaos} {\bf 1}, 49.}

\item[]{Danailov, M. B. \& Apai, P. [1994]
{"589 nm Generation by Intracavity Mixing in a Nd:YAG Laser,"}
{\it J. Appl. Phys.} {\bf 75}, 8240.}

\item[]{Falter S. et al. [1997] 
{"Dynamics and Stability of a Laser System with Second-order Nonlinearity,"}
{\it Opt. Lett.} {\bf 22}, 609.}

\item[]{Fan, T. Y. [1991]
{"Single-Axial Mode, Intracavity Doubled Nd:YAG Laser,"}
{\it IEEE J. Q. E.} {\bf EQ-27}, 2091.}

\item[]{James, G. E. et al. [1990]
{"Intermittency and Chaos in Intracavity Doubled Lasers. II,"}
{\it Phys. Rev. A} {\bf 41}, 2778.}

\item[]{James, G. E. et al. [1990]
{"Elimination of Chaos in an Intracavity-Doubled Nd:YAG laser,"}
{\it Opt. Lett.} {\bf 15}, 1141.}

\item[]{Kennedy, C. J. \& Barry, J. D. [1974]
{"Stability of an Intracavity Frequancy-Doubled Nd:YAG laser,"}
{\it IEEE J. Q.E} {\bf QE-10}, 596.}

\item[]{Liu, C. et al. [1997]
{"Influence of Noise on Chaotic Laser Dynamics,"}
{\it Phys. Rev. E} {\bf 55}, 6483.}

\item[]{Magni, V. et al. [1993]
{"Intracavity Frequency Doubling of a CW High-Power TEM00 Nd:YLF laser,"}
{\it Opt. Lett.} {\bf 18}, 2111.}

\item[]{Mandel, P. \& Wang, J. [1994]
{"Dynamic Independence of Pulsed Antiphased Modes,"}
{\it Opt. Lett.} {\bf 19}, 533.}

\item[]{Nagai, H. et al. [1992]
{"Low-Noise Operation of a Diode-Pumped Intracavity-Doubled
Nd:YAG Laser Using a Brewster Plate,"}
{\it IEEE J. Q. E.} {\bf QE-28}, 1165.}

\item[]{Otsuka, K. et al. [1997]
{"Clustering, Grouping, Self-induced Switching, and Controlled 
Dynamic Pattern Generation in a Antiphase Intracavity 
Second-harmonic Generation Laser,"}
{\it Phys. Rev. E} {\bf 56}, 4765. }

\item[]{Roy, R. et al. [1993]
{"Dynamics of a Multimode Laser with Nonlinear, Birefringent 
Intracavity Elements,"} in 
{\it Recent Developments in Quantum Optics}, ed. Inguva, R. 
(Plenum Press, New York). }

\item[]{Ustyugov, V. I. et al. [1997]
{"Self-Stabilization od a Continuous Wave Multimode Green 
Laser Due to Antiphase Dynamics,"}
{\it App. Ph. Lett.} {\bf 71}, 154.}

\item[]{Wang, J-Y. \& Mandel, P. [1993]
{"Antiphase Dyamics of Multimode Intracavity Second-Harmonic Generation,"}
{\it Phys. Rev. A} {\bf 48}, 671.}

\item[]{Wang, J-Y. at al. [1995]
{"Antiphase States in Intracavity Second Harmonic Generation,"}
{\it Quant. Semiclass. Opt.} {\bf 7}, 169.}

\item[]{Wiesenfeld, K. et al. [1990]
{"Observation of Antiphase States in a Multimode Laser,"}
{\it Phys. Rev. Lett.} {\bf 65}, 1749.}

\end{itemize}

\pagebreak

\begin{figure}
\caption{Schematic diagram of an intracavity-doubled solid-state laser. 
$L$ denotes the length of the cavity, $l$ is the length of the 
gain medium, $z_0$ is the distance of the gain medium from the cavity
mirror. }
\end{figure}

\begin{figure}
\caption{The evolution of the total intensity, 
$I_{tot} = \sum_{n=0}^{n=N}I(n,t)$, of the laser
for different values of the initial number of longitudinal 
modes: a) $N=3$, b) $N=70$, c) $N=80$, 
d) $N=5$, e) $N=100$, f) $N=250$ and different values of nonlinearity:
a), b), c) $\ep = 0.0001$, d), e), f) $\ep = 0.00012$; the case of 
the cross-saturation coefficient constant for all modes.}
\end{figure}

\begin{figure}
\caption{The dependence of the minimal 
number of modes, $N_{min}$ necessary to obtain a stabilization 
of the laser output as a function of the nonlinear coefficient, $\ep$; 
the comparison between 
the theoretical predictions (the straight line) and the results of
the numerical simulations; the case of 
the cross-saturation coefficient constant for all modes.}
\end{figure}

\begin{figure}
\caption{The evolution of the total intensity, 
$I_{tot} = \sum_{n=0}^{n=N}I(n,t)$, of the laser,
for different values of nonlinearity: 
a) $\ep=0.0001$, b) $\ep=0.00012$, c) $\ep=0.000125$, 
c) $\ep=0.00013$ and a constant value of the number of longitudinal 
modes, $N=100$; the case of the cross-saturation constant for all modes.}
\end{figure}

\begin{figure}
\caption{Different regions of the laser behavior, depending on the 
number of longitudinal modes, $N$, and the nonlinear coefficient, $\ep$.}
\end{figure}

\begin{figure}
\caption{The evolution of the total intensity, 
$I_{tot} = \sum_{n=0}^{n=N}I(n,t)$, of the laser
for different values of the initial 
number of longitudinal modes: a) $N=25$, b) $N=100$, c) $N=200$ and 
a constant value of nonlinearity, $\ep = 0.0001$; and also 
for different values of nonlinearity: 
a) $\ep= 1.0 10^{-6} $, b) $\ep= 1.0 10^{-5}$, c) $\sp=1.0 10^{-4} $ and
a constant number of longitudinal modes, $N=20$; the case of the
cross-saturation coefficient modulated by the effect of spatial 
hole-burning.}
\end{figure}

\pagebreak

\begin{figure}
\raisebox{-3.cm}[2cm][3.5cm]{
\psfig{figure=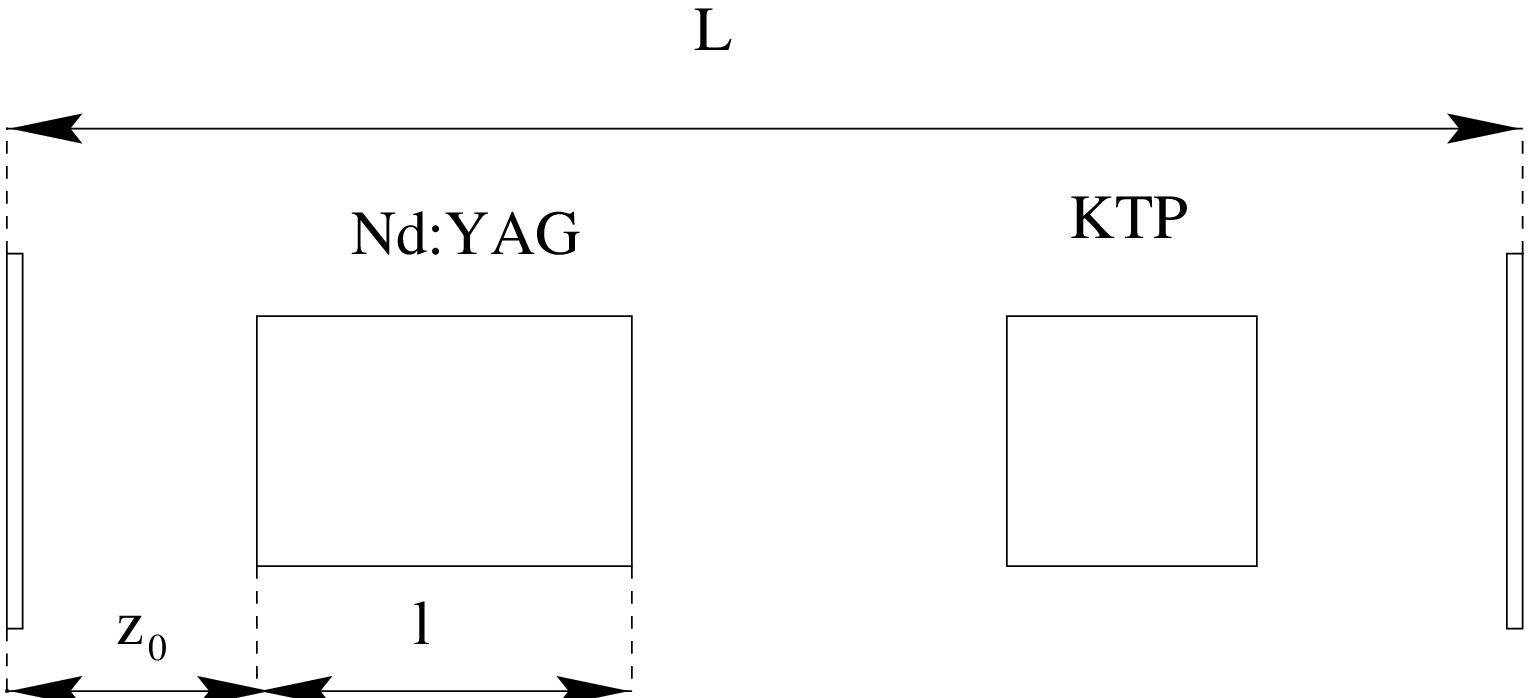,width=11cm}}
\label{input}
\end{figure}

\pagebreak

\begin{figure}
\raisebox{0.cm}[2cm][0cm]{
\hspace{-0.8cm}$I_{tot}$ \hspace*{8.6cm} $I_{tot}$}\\
\raisebox{-2.5cm}[1cm][-1cm]{
\hspace{-1.5cm}a)\hspace*{4.7cm}$t \, [ms] \hspace{4cm} d)
$\hspace{4.cm}$t\,[ms]$}\\
\raisebox{0cm}[2.0cm][4.0cm]{
\hspace*{-0.5cm}\psfig{figure=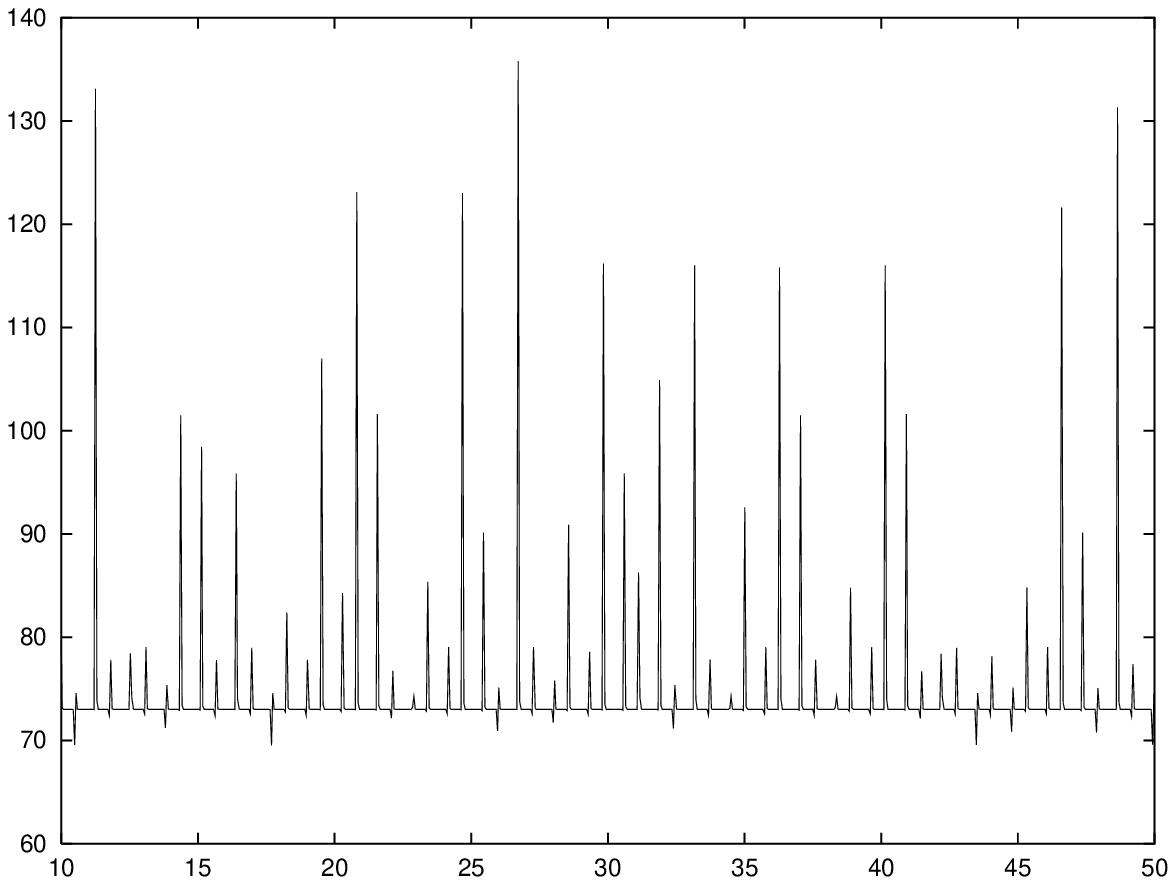,width=8.5cm}
\hspace*{1.0cm}\psfig{figure=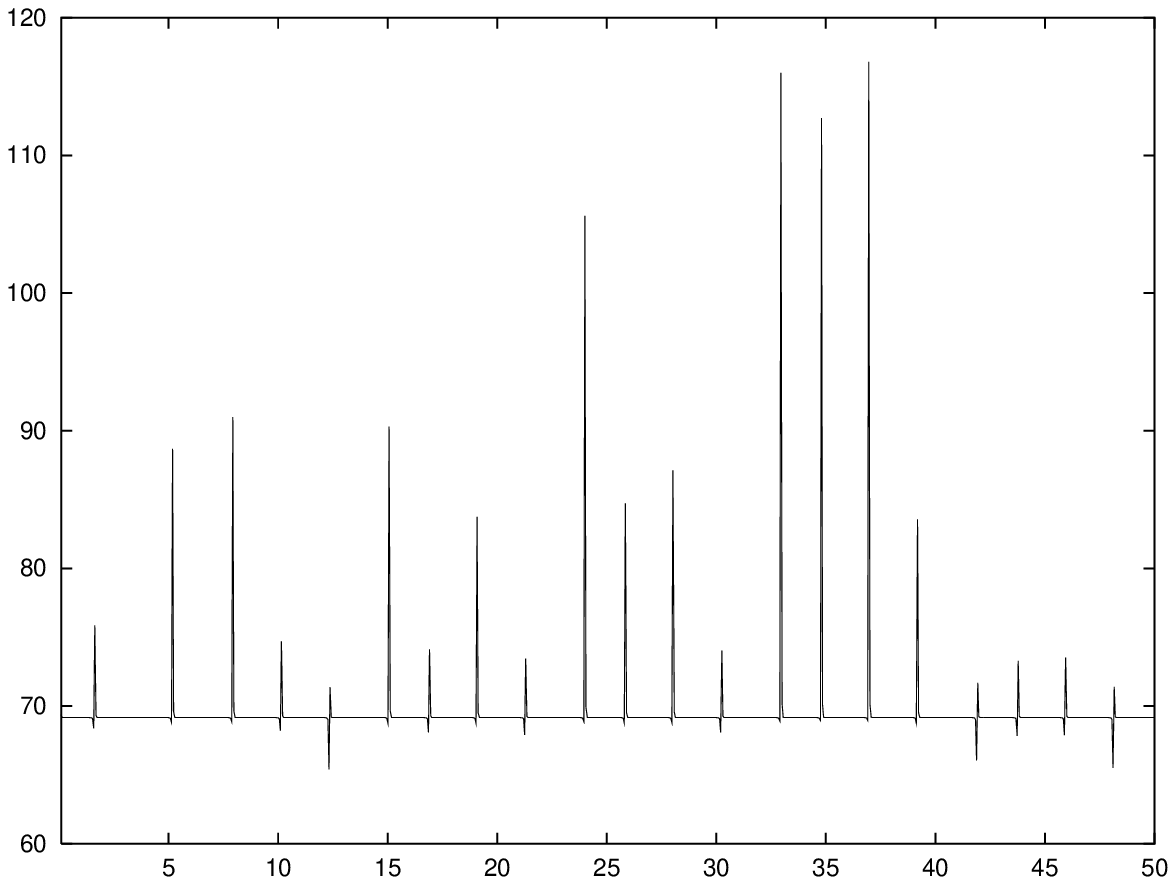,width=8.5cm}}\\
\raisebox{0.cm}[0cm][0cm]{
\hspace{-0.8cm}$I_{tot}$ \hspace*{8.6cm} $I_{tot}$}\\
\raisebox{-2.5cm}[1cm][-1cm]{
\hspace{-1.5cm}b)\hspace*{4.7cm}$t \, [ms] $\hspace{4cm} e)
\hspace{4.cm}$t\,[ms]$}\\
\raisebox{0cm}[2.0cm][4.0cm]{
\hspace*{-0.5cm}\psfig{figure=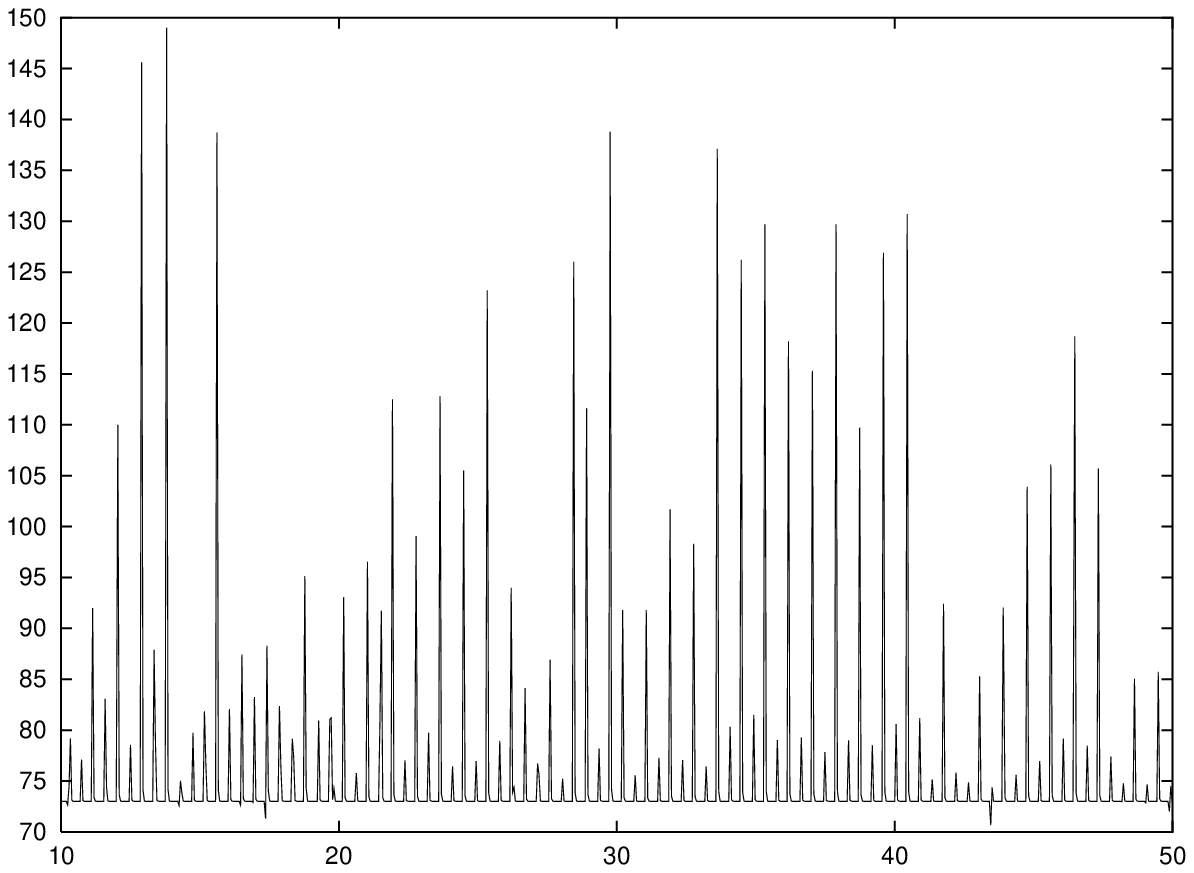,width=8.5cm}
\hspace*{1.0cm}\psfig{figure=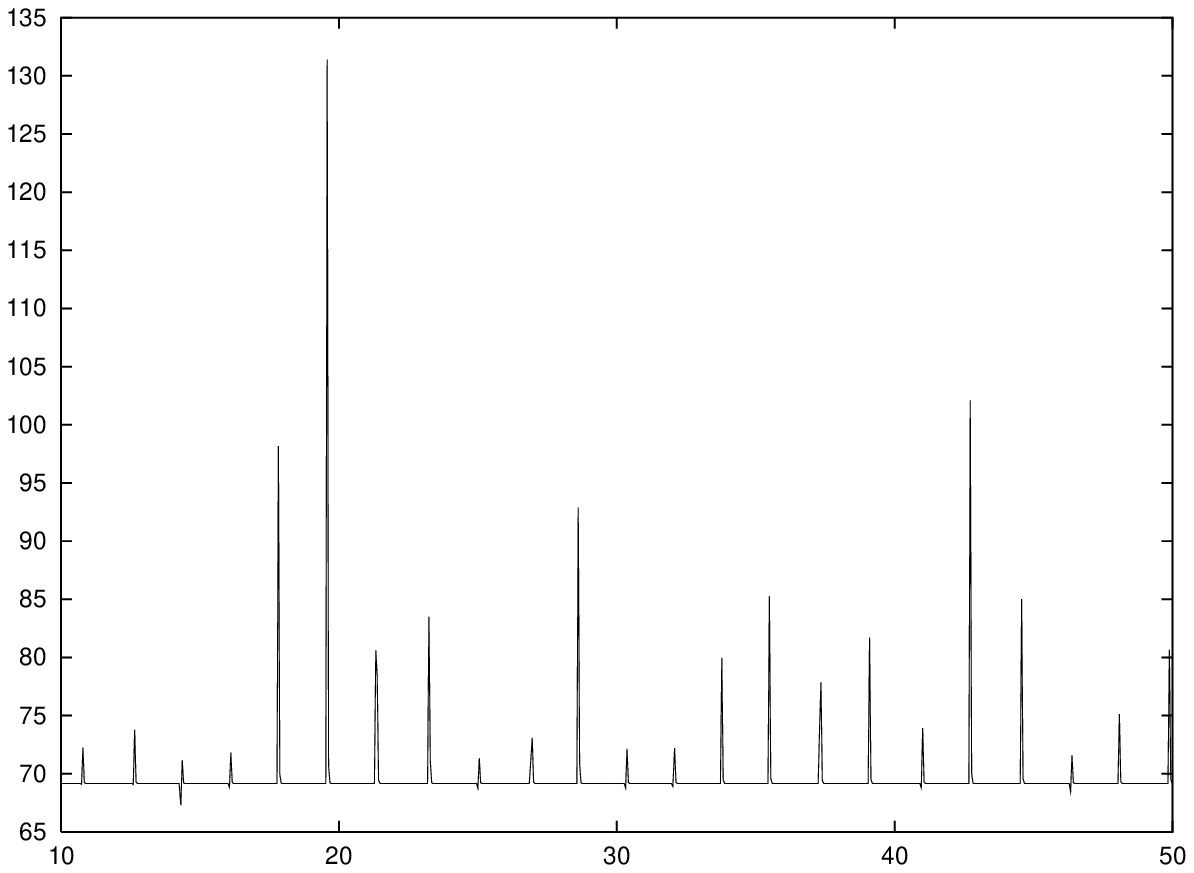,width=8.5cm}}
\raisebox{0.cm}[0cm][0cm]{
\hspace{-0.8cm}$I_{tot}$ \hspace*{8.6cm} $I_{tot}$}\\
\raisebox{-2.5cm}[1cm][-1cm]{
\hspace{-1.5cm}c)\hspace*{4.7cm}$t \, [ms] $  \hspace{4cm} f)
\hspace{4.cm}$t\,[ms]$}\\
\raisebox{0cm}[2.0cm][1.0cm]{
\hspace*{-0.5cm}\psfig{figure=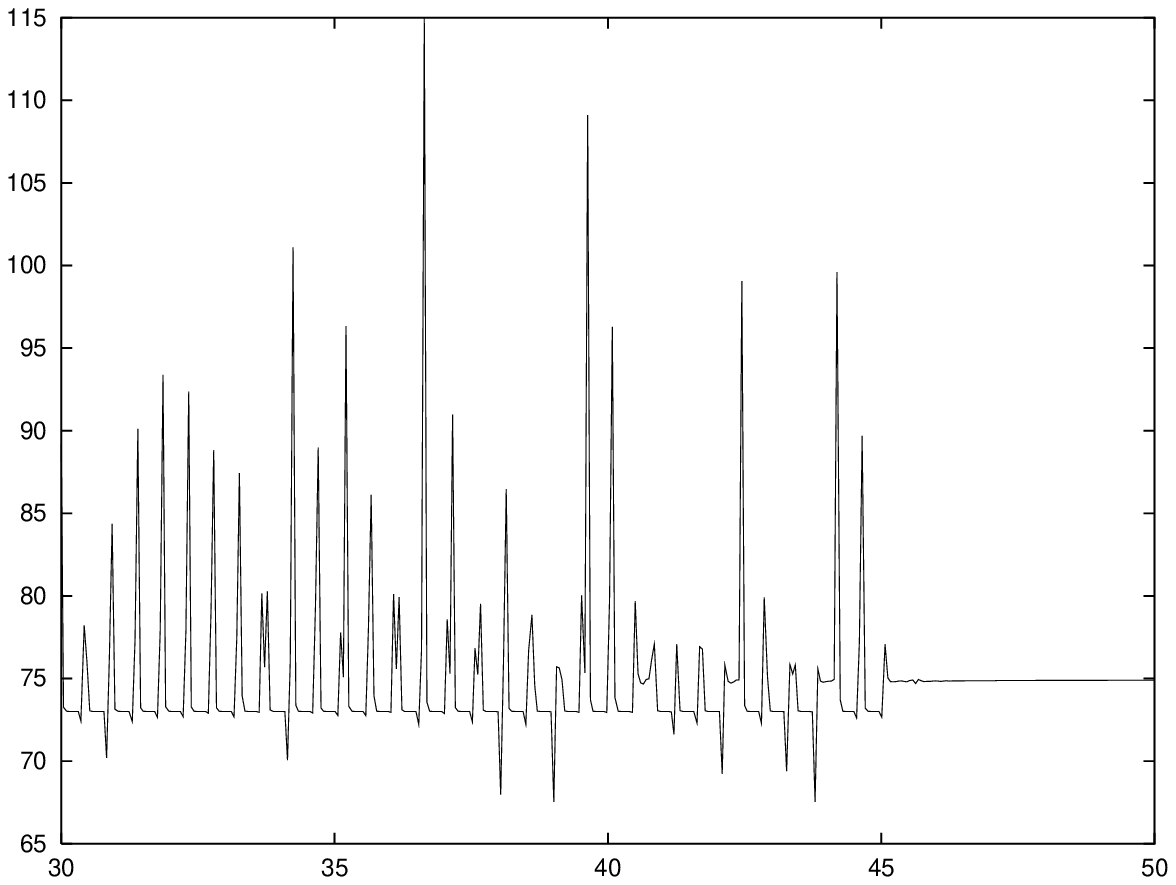,width=8.5cm}
\hspace*{1.0cm}\psfig{figure=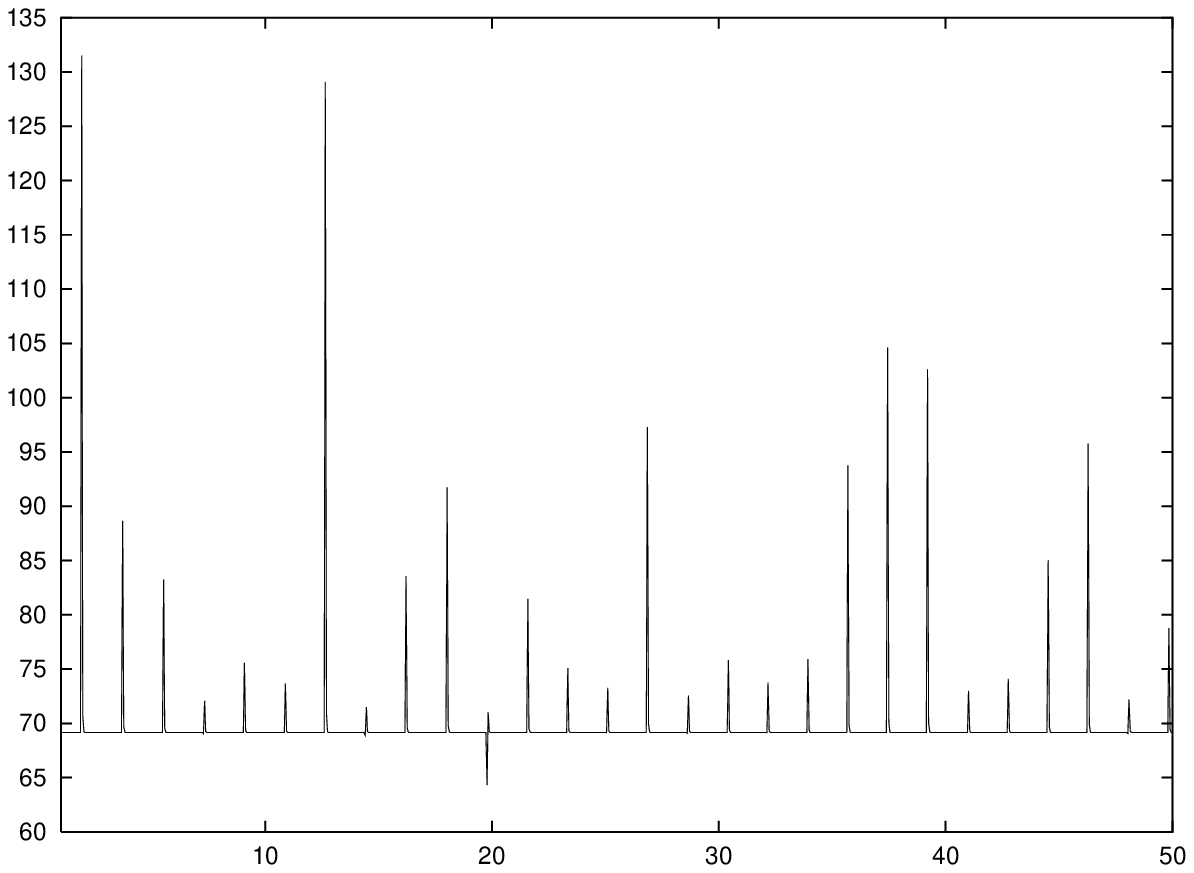,width=8.5cm}}
\label{fig2}
\end{figure}

\begin{figure}
\raisebox{-3cm}[1cm][-1cm]{
$\hspace{-0.5cm}N_{min}$}\\
\raisebox{-3.cm}[4cm][3.5cm]{
\hspace*{0.4cm}
\psfig{figure=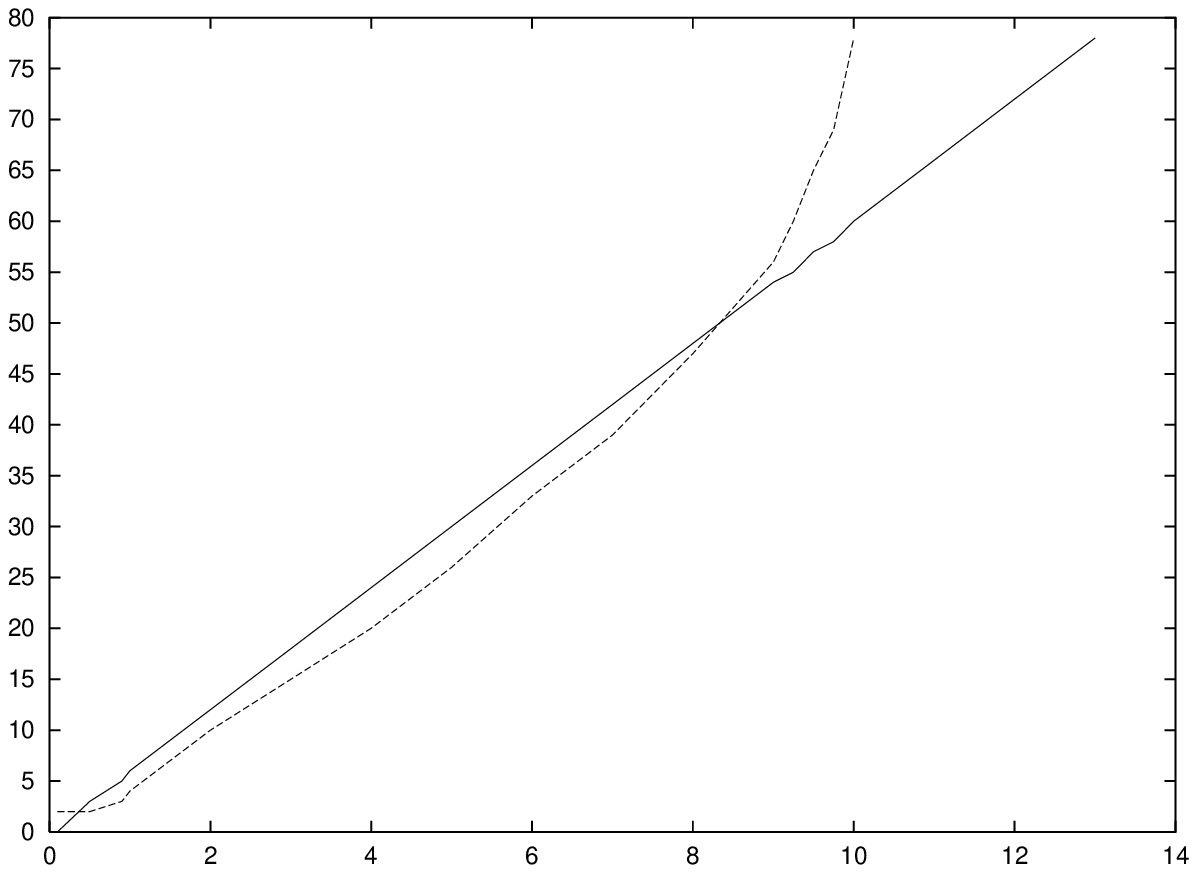,width=11cm}}
\raisebox{-3.7cm}[2cm][-2cm]{
$\hspace{-5.5cm}\ep \, [10^{-5}]$}\\
\label{fig6}
\end{figure}

\begin{figure}
\raisebox{0.cm}[2cm][0cm]{
\hspace{-0.8cm}$I_{tot}$ \hspace*{8.6cm} $I_{tot}$}\\
\raisebox{-2.5cm}[1cm][-1cm]{
\hspace{-1.5cm}a)\hspace*{4.7cm}$t \, [ms] \hspace{4cm} b)
$\hspace{4.cm}$t\,[ms]$}\\
\raisebox{0cm}[2.0cm][4.0cm]{
\hspace*{-0.5cm}\psfig{figure=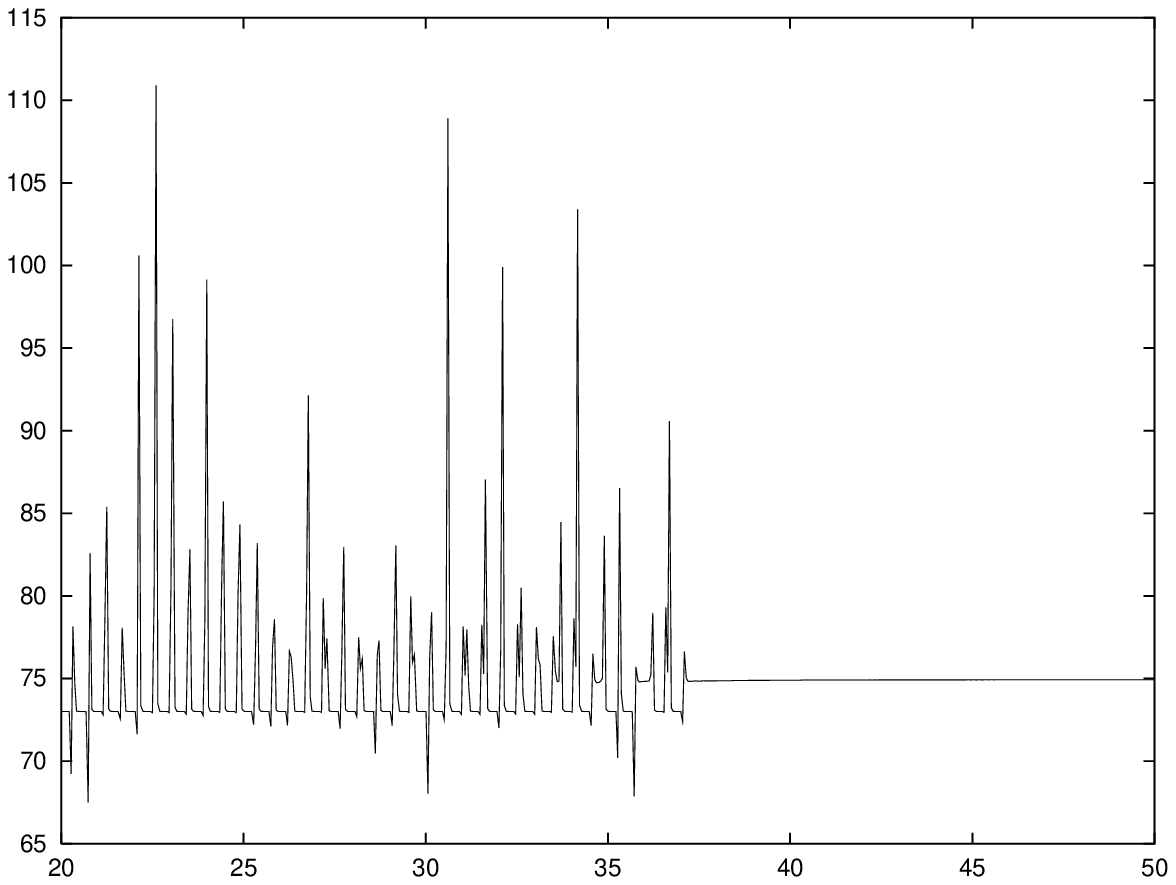,width=8.5cm}
\hspace*{1.0cm}\psfig{figure=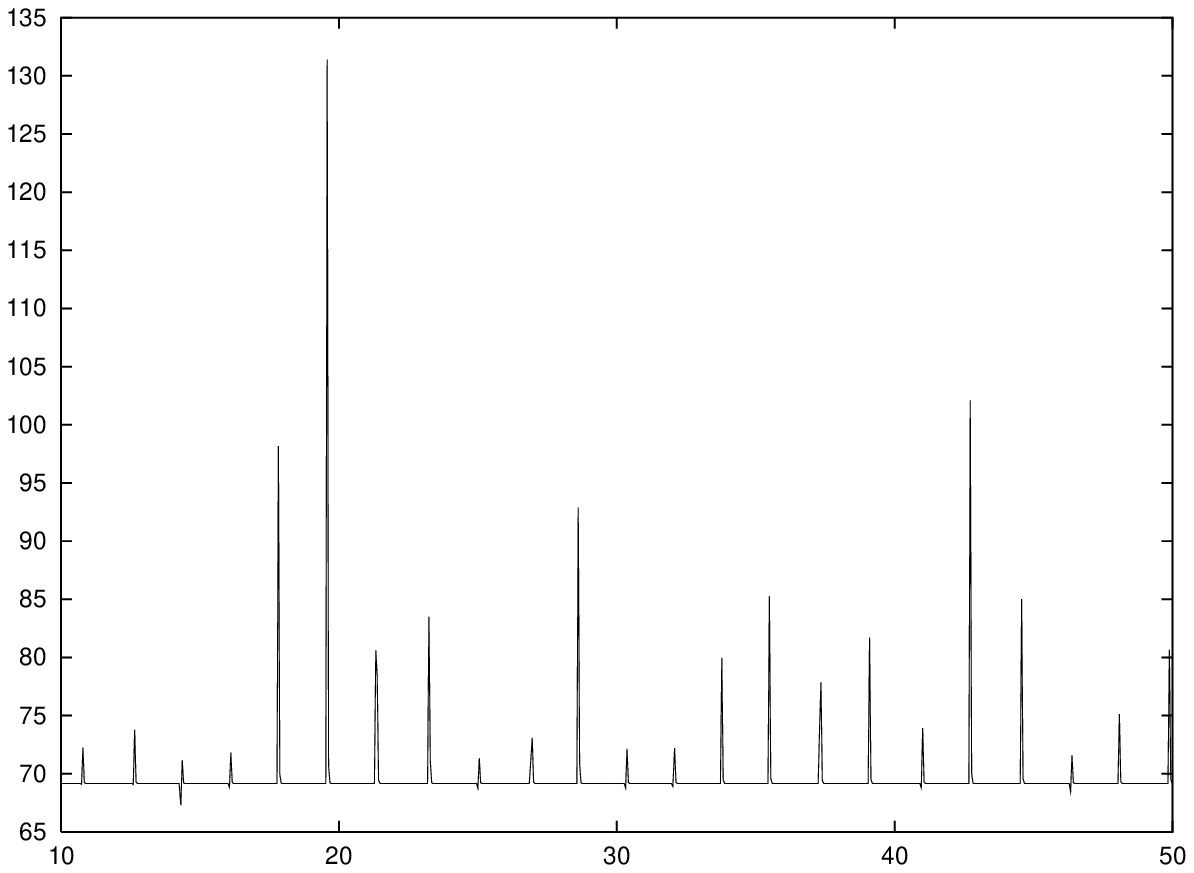,width=8.5cm}}\\
\raisebox{0.cm}[0cm][0cm]{
\hspace{-0.8cm}$I_{tot}$ \hspace*{8.6cm} $I_{tot}$}\\
\raisebox{-2.5cm}[1cm][-1cm]{
\hspace{-1.5cm}c)\hspace*{4.7cm}$t \, [ms] $\hspace{4cm} d)
\hspace{4.cm}$t\,[ms]$}\\
\raisebox{0cm}[2.0cm][4.0cm]{
\hspace*{-0.5cm}\psfig{figure=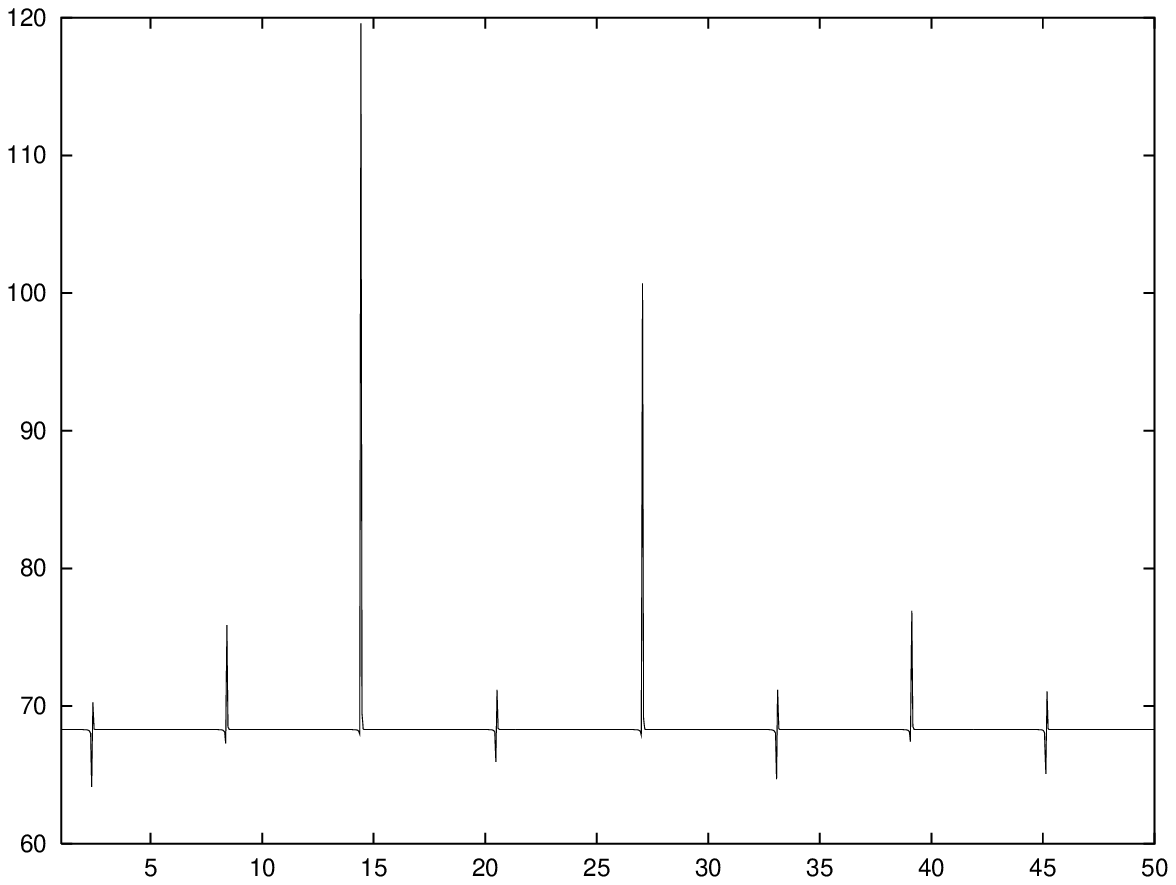,width=8.5cm}
\hspace*{1.0cm}\psfig{figure=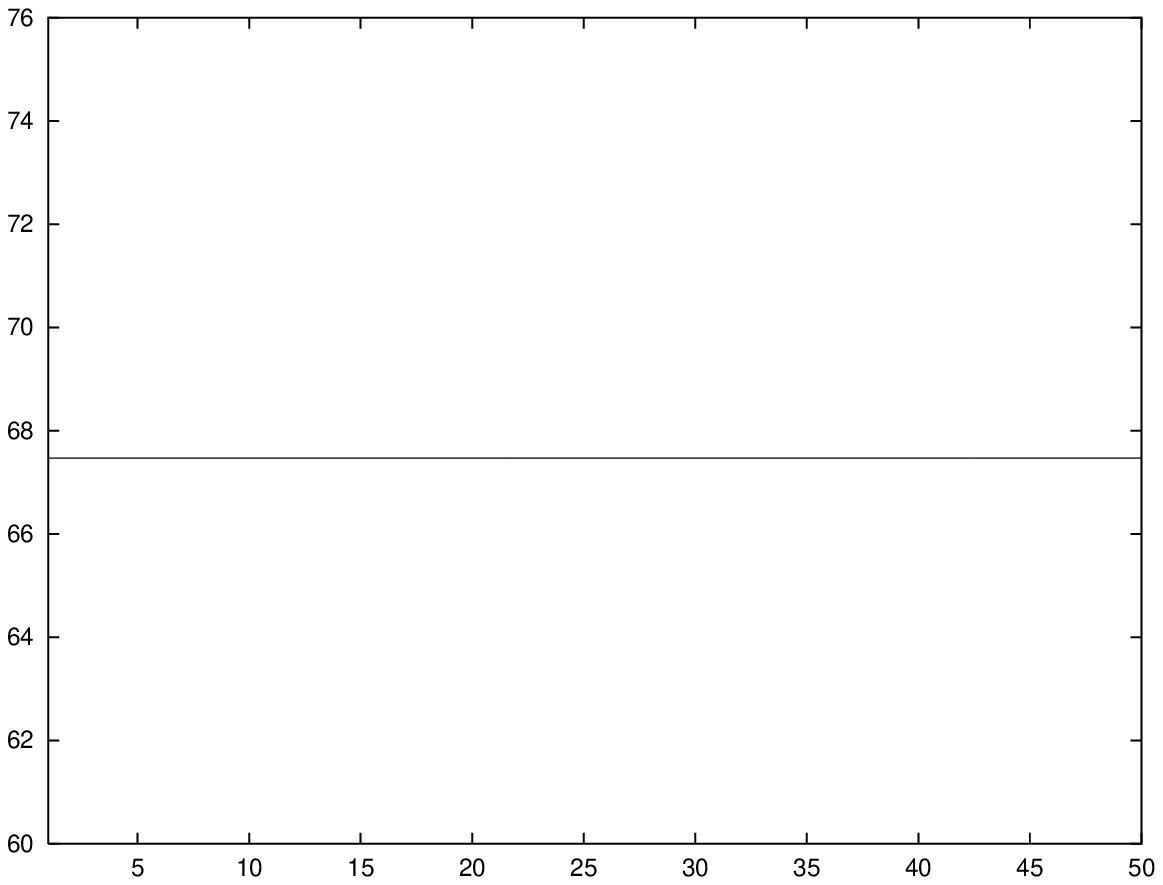,width=8.5cm}}
\label{fig8}
\end{figure}

\begin{figure}
\raisebox{-6.5cm}[1cm][-1cm]{
$N$}\\
\raisebox{-3.cm}[7cm][3.5cm]{
\hspace{3mm}\psfig{figure=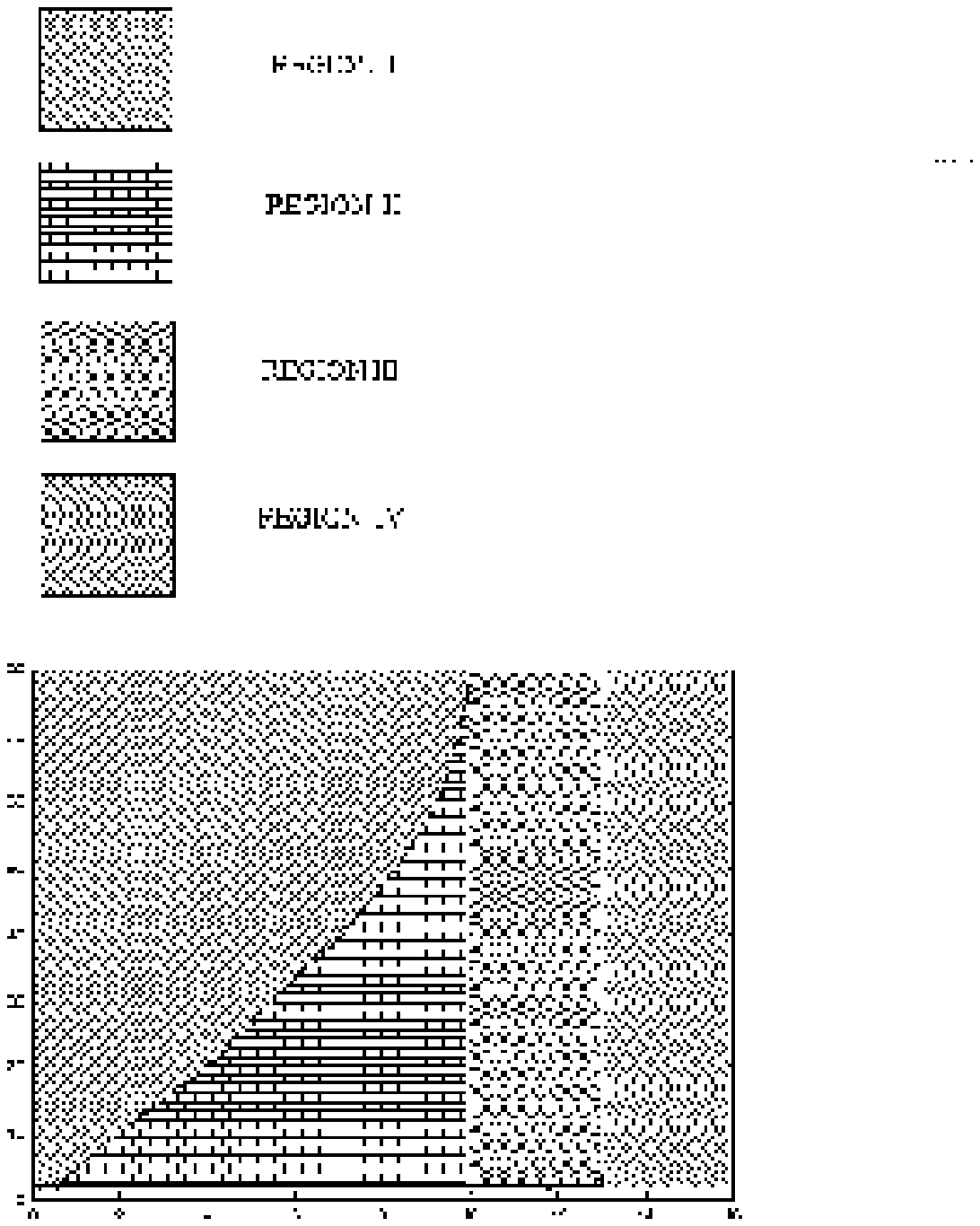,width=12cm}}
\raisebox{-3.0cm}[2cm][-2cm]{
$\hspace{-8.0cm}\ep \, [10^{-5}]$}\\
\label{fig9}
\end{figure}

\begin{figure}
\raisebox{0.cm}[2cm][0cm]{
\hspace{-0.8cm}$I_{tot}$ \hspace*{8.6cm} $I_{tot}$}\\
\raisebox{-2.5cm}[1cm][-1cm]{
\hspace{-1.5cm}a)\hspace*{4.7cm}$t \, [ms] \hspace{4cm} d)
$\hspace{4.cm}$t\,[ms]$}\\
\raisebox{0cm}[2.0cm][4.0cm]{
\hspace*{-0.5cm}\psfig{figure=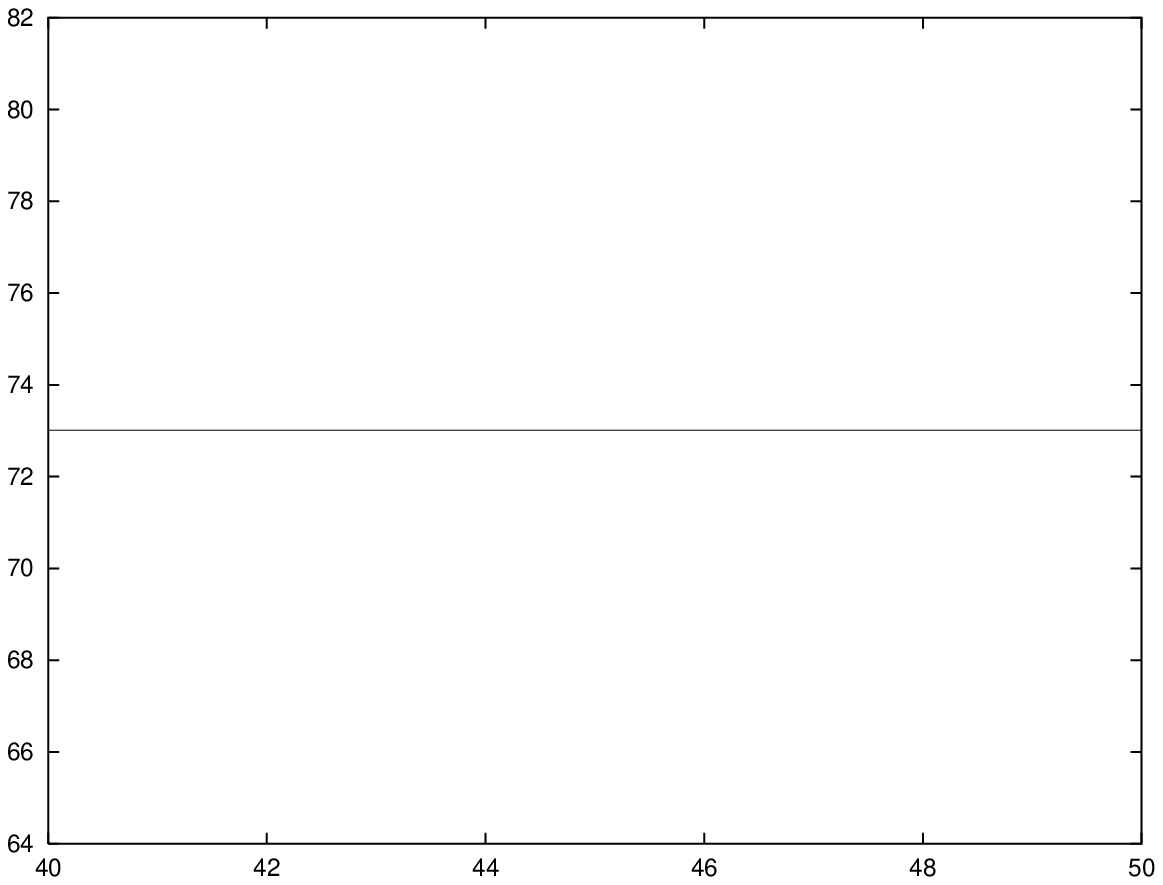,width=8.5cm}
\hspace*{1.0cm}\psfig{figure=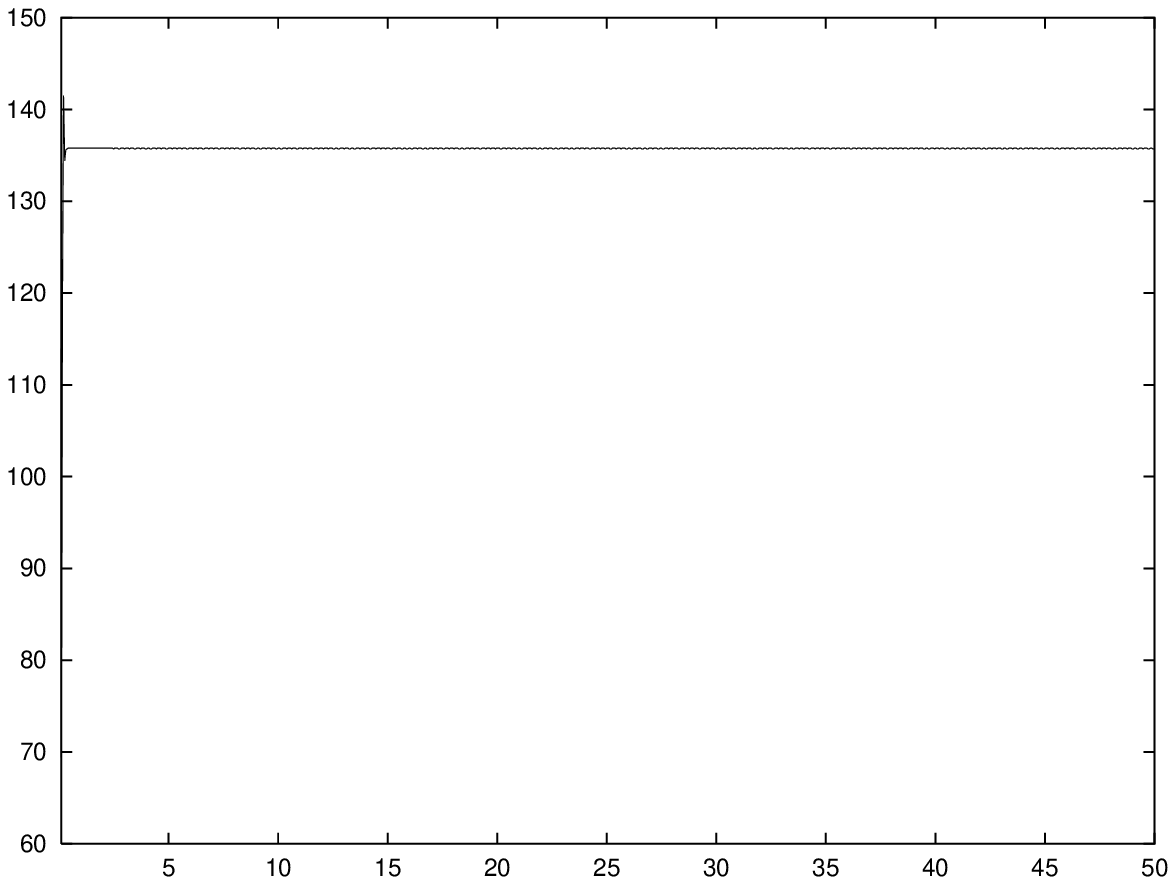,width=8.5cm}}\\
\raisebox{0.cm}[0cm][0cm]{
\hspace{-0.8cm}$I_{tot}$ \hspace*{8.6cm} $I_{tot}$}\\
\raisebox{-2.5cm}[1cm][-1cm]{
\hspace{-1.5cm}b)\hspace*{4.7cm}$t \, [ms] \hspace{4cm} e)
$\hspace{4.cm}$t\,[ms]$}\\
\raisebox{0cm}[2.0cm][4.0cm]{
\hspace*{-0.5cm}\psfig{figure=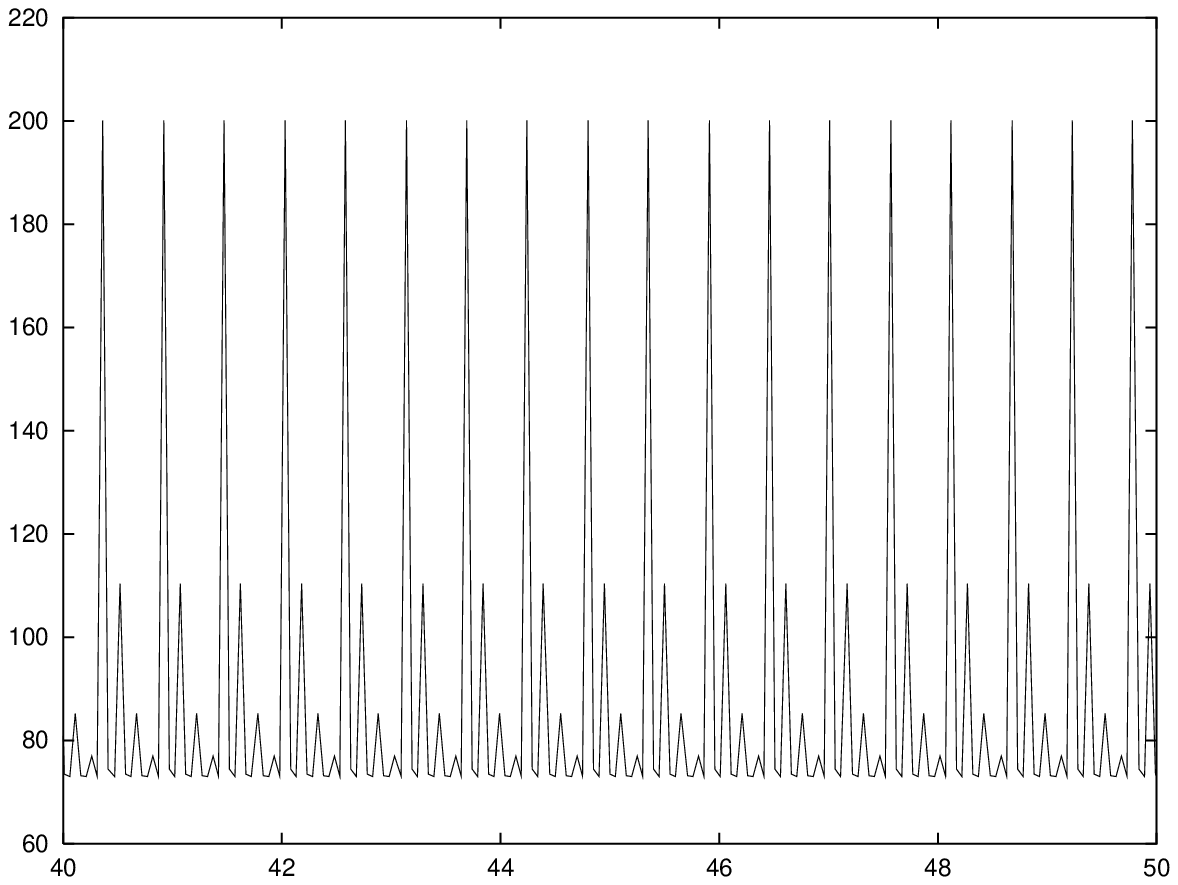,width=8.5cm}
\hspace*{1.0cm}\psfig{figure=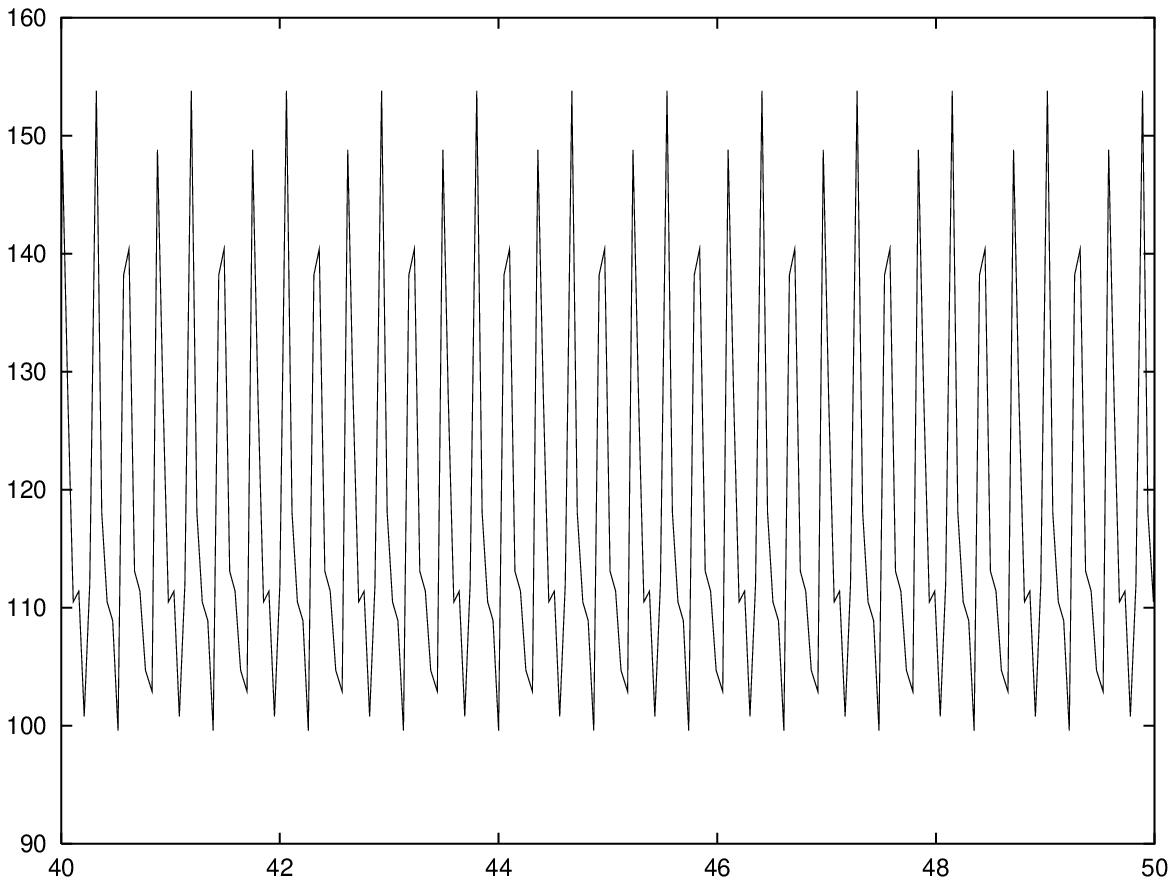,width=8.5cm}}
\raisebox{0.cm}[0cm][0cm]{
\hspace{-0.8cm}$I_{tot}$ \hspace*{8.6cm} $I_{tot}$}\\
\raisebox{-2.5cm}[1cm][-1cm]{
\hspace{-1.5cm}c)\hspace*{4.7cm}$t \, [ms] \hspace{4cm} f)
$\hspace{4.cm}$t\,[ms]$}\\
\raisebox{0cm}[2.0cm][1.0cm]{
\hspace*{-0.5cm}\psfig{figure=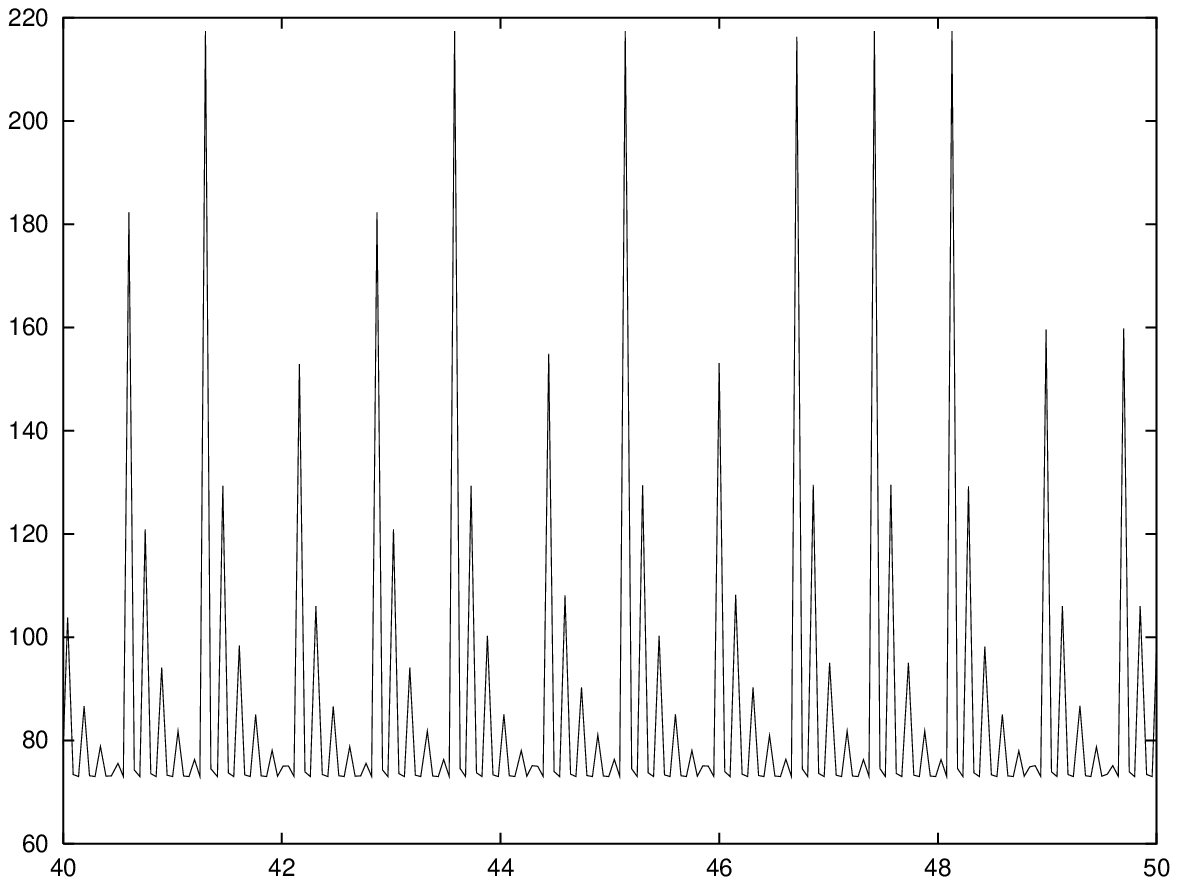,width=8.5cm}
\hspace*{1.0cm}\psfig{figure=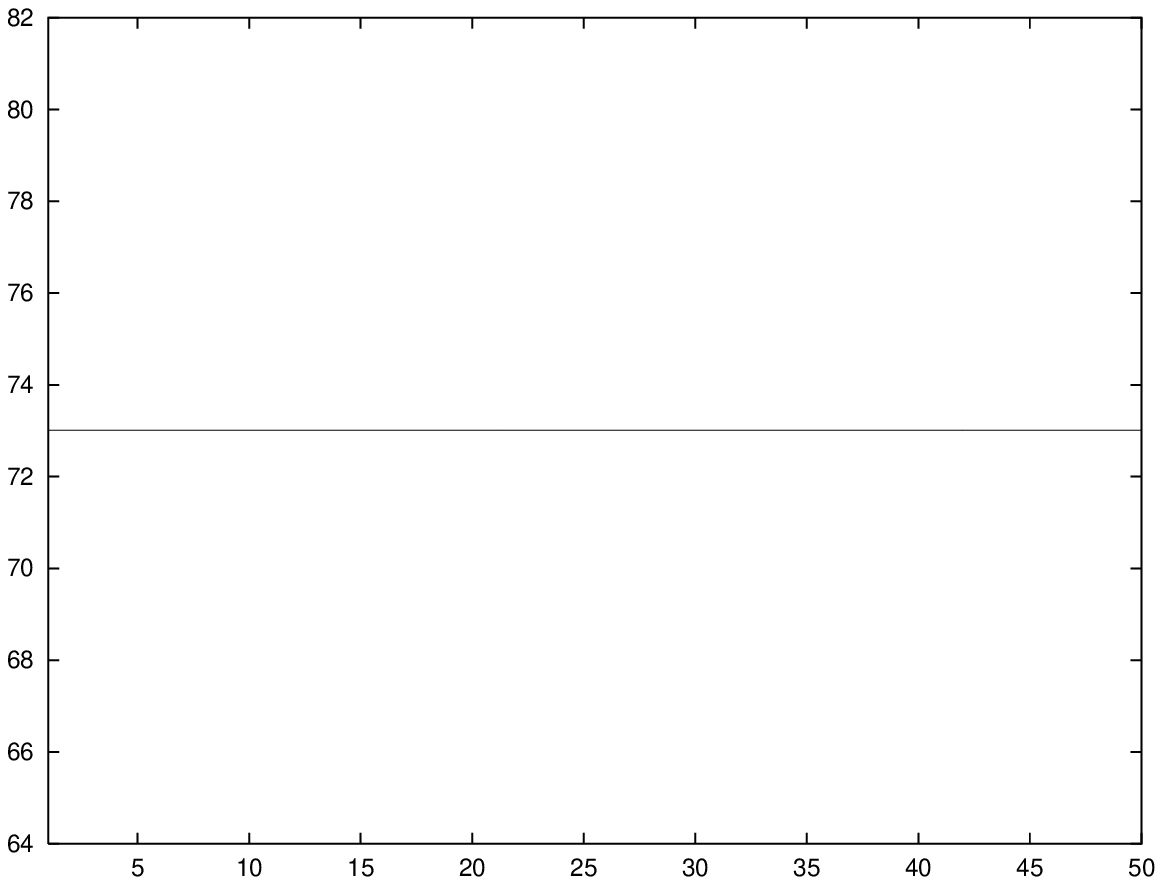,width=8.5cm}}
\label{fig11}
\end{figure}

\end{document}